# An Updated Lagrangian Particle Hydrodynamics (ULPH)-NOSBPD Coupling Approach for Modeling Fluid-Structure Interaction Problems

Zhen Wang[a], Junsong Xiong[b], Xiang Liu[b], Lisheng Liu[b], Xin Lai*,[a]

*[a] Hubei Key Laboratory of Theory and Application of Advanced Materials Mechanics, Wuhan University of Technology, Wuhan 430070, China*
*[b] Department of Engineering Structure and Mechanics, Wuhan University of Technology, Wuhan 430070, China*

*corresponding author: laixin@whut.edu.cn

**Abstract**

In order to solve the fluid-structure interaction problem of Newtonian fluid, a fluid-structure interaction approach is proposed based on Non-ordinary State-based Peridynamics (NOSB-PD) and Updated Lagrangian particle Hydrodynamics (ULPH), to simulate the fluid-structure interaction problem in which large geometric deformation and material failure are considered. In the coupled framework, the NOSB-PD theory is used to describe the deformation and fracture of the solid material structure. ULPH is applied to describe the flow of Newtonian fluids due to its advantages in computational accuracy. The framework effectively utilizes the advantages of NOSB-PD theory for solving discontinuous problems and ULPH theory for solving fluid problems and has good computational stability and robustness. To deal with the interface of fluid structure, a fluid-structure coupling algorithm using pressure as the transmission medium is established. The dynamic model of solid structure and the PD-ULPH fluid-structure interaction model involving large deformation are verified by several numerical validations, which are in good agreement with the analytical solution, the available experimental data and other numerical results, that demonstrates the accuracy and effectiveness of the proposed method in solving the fluid-structure interaction problem. Overall, the fluid-structure interaction model based on ULPH and NOSB-PD established in this paper provides a new idea for the numerical solution of fluid-structure interaction and a promising approach for engineering design and experimental prediction.

## 1. Introduction

Facing the strategic needs of maritime power, ships and ocean engineering have been vigorously developed, and the related fields of structural mechanics and hydrodynamics have yielded fruitful results[1][2]. Among them, typical high-speed hydrodynamic problems such as high-speed motion of vehicles in water, underwater explosion and structural damage[3] are closely related to the comprehensive performance of modern ships and their weapons and equipment. However, these problems are multi-physical field problems, and all involve the state of interaction between moving or deformed structures and surrounding or internal fluids. They all belong to Fluid-Structure Interaction (FSI) problems.

Fluid-structure interaction problems are often complex and involve many nonlinear factors, and it is difficult to obtain analytical solutions through theoretical derivation. Numerical simulation and model test have become two common ways to analyze fluid-structure interaction problems. For model test, the repeatability of test data is poor, and the cost of trial and error is high. In contrast, numerical simulation has the advantages of low cost, short cycle and clear physical process, and plays an increasingly important role in the field of ship and ocean engineering.

In recent years, with the development of computer technology, many numerical calculation methods have emerged. Currently, based on different discretization and solution forms, numerical simulation methods can be classified into grid methods described by Euler, such as Volume method (VOF) [4][5], level set method (LS) [6][7], Lattice Boltzmann method (LBM) [8][9] and forward tracking method [10][11]. And Lagrangian description of meshless methods such as smooth particle Hydrodynamics (SPH)[12], reproducing kernel particle method (RKPM)[13], moving particle Semi-implicit (MPS)[14], and material point method (MPM)[15]. And, more recently, the updated Lagrangian Particle Hydrodynamics (ULPH) originally proposed by Tu and Li[16]. The mesh method is easy to suffer from mesh distortion when solving the fluid-structure coupling problem with large deformation. The latter meshless method benefits from natural Lagrangian characteristics and gradually complete particle approximation theory advantages and is not limited by boundary deformation when simulating large deformation problems[17], so it has been widely used in fluid-structure interaction problems.

ULPH, is another meshless particle method that has been successfully implemented to solve fluid dynamics problems. Inspired by near-field dynamics[18], RKPM[13] and SPH, the local differential operator in the Navier-Stokes equations is replaced by a nonlocal differential operator (NDO). Since the nonlocal continuous function space contains a function space much larger than the local continuous function space, It may capture more physical content than the local continuum CFD method, and compared with the classical SPH method, it eliminates the tensile instability and the accuracy loss caused by the kernel approximation, so it is more suitable for describing the fluid motion in more complex flow fields. ULPH uses the updated Lagrangian formulation, and selects the current configuration as the updated reference configuration instead of using the initial configuration as the reference configuration, and continuously updates the reference configuration while the calculation is carried out. Instead of using the variables as SPH and molecular dynamics use the total Lagrangian method are defined and described in the initial configuration, or the initial configuration is chosen as the reference configuration. As pointed out in [19], the use of the updated configuration as a reference configuration is advantageous as another rapidly developing meshless method. The basic idea of near-field dynamics (PD), proposed by SILLING[20], is a nonlocal continuous theory, which uses nonlocal integral equations, and is a non-local integral equation. Discontinuous displacement field can be naturally included in the governing equations, so it has natural advantages in simulating crack initiation and propagation in materials, and has become a research hotspot. Near field dynamics, there are two different branches, based on the key of the near field dynamics (bond -based Peridynamics, BBPD)[21] and unconventional state of near field dynamics (state-based Peridynamics, NOSB - PD) [22]. At present, a complete discretization model and numerical integration algorithm for the ULPH and BBPD coupling formulas have been successfully established. And it is used to simulate ice-water interaction under impact load [23], which is both ice fragmentation and fracture under impact load. Many Lagrangian flow problems, such as the fluid-structure interaction problem [24]. Yan et al.[25][26] developed a set of high-order nonlocal differential operators in the ULPH framework and applied them to solve multiphase flow problems. The results show that the ULPH method has better accuracy than SPH in modeling and simulation of multiphase flow problems.

Despite the success in the problem of coupling BBPD with ULPH, the interaction between two material points in the bond-based near-field dynamics depends only on the deformation of the bond between that material point. This assumption restricts the Poisson ratio of the solid model, so solving typical FSI-based problems by coupling PD with ULPH is still a difficult task. Since ULPH-NOSBPD has a good potential for structural analysis, combining their advantages is an important research topic. The motivation for this work is to develop an ULPH-NOSBPD method that is able to handle complex fluid flows and large structural deformations, even failures, simultaneously.

## 2. Numerical approach

### 2.1. Update Lagrangian particle hydrodynamics for Newtonian fluids

#### 2.1.1. Governing equations

In the present work, the fluid is assumed weakly compressible without consideration of thermal effects. When the pressure peak of the weakly compressible fluid is lower than 1GPa, the energy changes have little effect on the fluid features. The fluid is considered as isentropic. The fluid dynamics formulation can be solved using the Navier-Stokes equations. The Navier-Stokes equations can effectively describe the relationship between the velocity, pressure, density and temperature of a fluid. It is a set of coupled differential equations that can theoretically be solved using methods from calculus for a given flow problem.

In the Lagrangian form, the governing equations for the isentropic fluid consist of mass and momentum conservation laws, and the general form of the governing equations[25] are written as follows:

$$\frac{D\rho}{Dt} = -\rho \nabla \cdot \mathbf{v} \tag{2-1}$$

$$\frac{D\mathbf{v}}{Dt} = \frac{1}{\rho} \nabla \cdot \sigma + \mathbf{g} \tag{2-2}$$

where $D/Dt$ represents the material time derivative, $\mathbf{v}$ represents the velocity vector, $\rho$ is the fluid density, g is the gravity acceleration and 9.81m/s$^2$ is used in the present paper.

σ represents the Cauchy stress tensor[25], which is the summation of a pressure term $-p\mathbf{I}$ (hydrostatic stress or volumetric stress) and a viscosity term $\tau$ (deviatoric stress) as

$$\sigma = -p\mathbf{I} + \tau \tag{2-3}$$

where $\mathbf{I}$ is the unit second-order tensor. The viscous stress $\tau$ is expressed as

$$\tau = 2\mu\dot{\varepsilon} \tag{2-4}$$

where $\mu$ is the dynamic viscosity, $\dot{\varepsilon}$ is the rate of shear strain tensor as follows:

$$\dot{\varepsilon} = \frac{1}{2}\left[\nabla \otimes \mathbf{v} + (\nabla \otimes \mathbf{v})^T\right] - \frac{1}{3}(\nabla \cdot \mathbf{v})\mathbf{I} \tag{2-5}$$

The above Navier-Stokes equations are non-closed at this stage, Additional equations of state need to be added to establish the connection between pressure p and density $\rho$. For the weakly compressible fluid, the Tait equation[27] can be connected to solve the Navier-Stokes equations. In the paper, following a linearized form of Tait equation as the equation of state, the evolution of the pressure from the density is determined as

$$p = \frac{c_0^2 \rho_0}{\gamma}\left[\left(\frac{\rho}{\rho_0}\right)^\gamma - 1\right] \tag{2-6}$$

where $\rho_0$ is the density in the reference configuration and $\gamma$ Is the characteristic index coefficient, for water $\gamma$ is usually 7. $c_0$ is the reference sound speed to control the compressibility of the fluid, which should satisfy the density variation of less than 1%. In order to satisfy the weakly compressible properties, the reference sound speed [26] should be taken as follows.

$$c_0 \geq 10\max\left(\sqrt{p_{max}/\rho_0}, U_{max}\right) \tag{2-7}$$

where $p_{max}$ and $U_{max}$ represent the maximum expected pressure and velocity in the computational domain, respectively. The true sound speed of the fluid is not used in order to increase the time step size and thus improve the computational efficiency.

### 2.1.2. Optimal nonlocal differential operators

The calculation of gradient and divergence in flow field is a key part of computational fluid dynamics. Because the continuity equation is solved through the divergence of the velocities. However, the momentum equation is solved by the divergence of the stress and the gradient of the velocity. Equations (2-1)~(2-3) are the governing equations in local form, and the grid-like numerical methods divide the entire computational domain into grids and apply local theory ideas to solve the governing equations numerically. For the meshless method, the whole computational domain needs to be discretized into particles, each particle has the properties of density, mass, pressure, velocity and other physical quantities, and then the governing equations are discretized by using the idea of non-local theory.

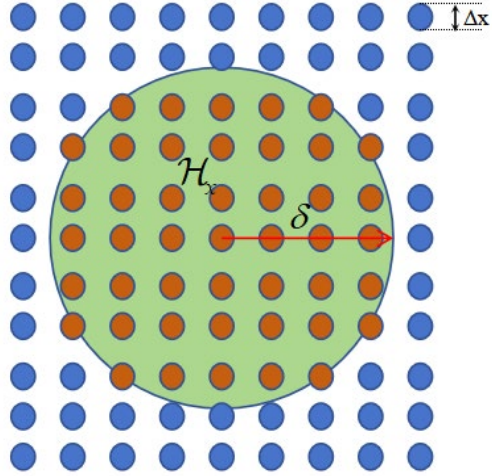

Fig. 1.Nonlocal theoretical models

In the ULPH framework[16], the nonlocal differential operator is used instead of the local operator to calculate such as the divergence, gradient and curl.

$$\nabla_I \cdot (\bullet) := \int_{\mathcal{H}_I} \omega(x_{IJ})(\Delta(\bullet)) \cdot \left(\mathbf{M}_I^{-1} x_{IJ}\right) dV_J, \tag{2-8}$$

$$\nabla_I \otimes (\bullet) := \int_{\mathcal{H}_I} \omega(x_{IJ})(\Delta(\bullet)) \otimes \left(\mathbf{M}_I^{-1} x_{IJ}\right) dV_J \tag{2-9}$$

$$\nabla_I \times (\bullet) := \int_{\mathcal{H}_I} \omega(x_{IJ})\left(\mathbf{M}_I^{-1} x_{IJ}\right) \times (\Delta(\bullet)) dV_J \tag{2-10}$$

where the operators ($\nabla \cdot$), ($\nabla \otimes$), ($\nabla \times$) represent the nonlocal divergence, gradient and curl operators, respectively; $\mathcal{H}_I$ represents the support domain of particle I, as shown in Figure 1, the subscript J denotes the family member of particle I, and $x_{IJ} = x_J - x_I$. The symbol $(\bullet)$ represents the arbitrary field, and $\Delta(\bullet) := (\bullet)_J - (\bullet)_I$ is the finite difference operator. $\omega(x_{IJ})$ is the kernel function and must meet specific criteria, more on that later. M is the shape tensor defined in the current configuration or the updated configuration[16].

$$\mathbf{M}_I := \int_{\mathcal{H}_I} \omega(x_{IJ}) x_{IJ} \otimes x_{IJ} dV_J \tag{2-11}$$

The computational domain of ULPH is discretized as a sequence of particles with physical properties, so that, according to Eq. 21, the above nonlocal differential operator can be discretely rewritten as

$$\nabla_I \cdot (\bullet) = \sum_{J=1}^{N} \omega(x_{IJ})(\Delta(\bullet)) \cdot \left(\mathbf{M}_I^{-1} x_{IJ}\right) V_J \tag{2-12}$$

$$\nabla_I \otimes (\bullet) = \sum_{J=1}^{N} \omega(x_{IJ})(\Delta(\bullet)) \otimes \left(\mathbf{M}_I^{-1} x_{IJ}\right) V_J \tag{2-13}$$

$$\nabla_I \times (\bullet) = \sum_{J=1}^{N} \omega(x_{IJ})\left(\mathbf{M}_I^{-1} x_{IJ}\right) \times (\Delta(\bullet)) V_J \tag{2-14}$$

The discretized form of the moment matrix Eq. (24)is reformulated as

$$\mathbf{M}_I = \sum_{J=1}^{N} \omega(x_{IJ}) x_{IJ} \otimes x_{IJ} V_J \qquad (2\text{-}15)$$

### 2.1.3. Discrete form of governing equations

In ULPH, we choose current configuration $\Omega_n$ at time t = tn as the reference configuration as described in [16][28], This is shown in Figure 1,and keep updating the referential configuration as the computation proceeds. Therefore, the governing equations below are calculated under the current configuration

Based on the above nonlocal differential operators and Peridynamic theory [28],By substituting the nonlocal divergence operators (2-12) and (2-13) into the continuous equation (2-1), we can obtain the continuous equation in the nonlocal discrete form in the ULPH framework

$$\frac{D\rho_I}{Dt} = -\rho_I \sum_{J=1}^{N} \omega(x_{IJ})(v_J - v_I) \mathbf{M}_I^{-1} x_{IJ} V_J \qquad (2\text{-}16)$$

where $V_J$ is the volume of the fluid particle, $V_J = m_J / \rho_J$, m is the mass of the fluid particle. $\omega(x_{IJ})$ is the kernel function,The selection of kernel function is related to the accuracy, efficiency and stability of numerical simulation.In this paper, the Gaussian kernel function [9] is adopted to all the simulations as the influence function ω, and it is defined as

$$\omega(r,h) = \alpha_d \left( e^{-(r/h)^2} - C \right) \quad r \leq \delta \qquad (2\text{-}17)$$

where r is the distance between two neighboring particles I and J, h is the smoothing length typically set to $h = 1.2\Delta x$, Where $\Delta x$ is the initial spacing of the particles, $\delta(\delta = 3h)$ is the size of the support domain of the particle, d is the spatial dimension and C is generally set to $e^{-9}$, and Normalized coefficients $\alpha_d$ is a coefficient associated to spatial dimension and the smoothing length, as the following form,

$$\alpha_d = \frac{1}{h^d \pi^{d/2} \left( 1 - 10e^{-9} \right)} \qquad (2\text{-}18)$$

According to the theory of state-based Peridynamic theory, the nonlocal form of the momentum equation [22] can be defined as

$$\rho_I \frac{D\mathbf{v}_I}{Dt} = \int_{\mathcal{H}_I} \left( \mathbf{T}_I(\mathbf{x}_{IJ}) - \mathbf{T}_J(\mathbf{x}_{JI}) \right) dV_J + \mathbf{b}_I \qquad (2\text{-}19)$$

$\mathbf{T}_I(\mathbf{x}_{IJ})$ and $\mathbf{T}_J(\mathbf{x}_{JI})$ denote the force state vector acting on particle I and particle J, $\mathbf{b}_I$ is the external body force.

$$\mathbf{T}_I(\mathbf{x}_{IJ}) = \omega(\mathbf{x}_{IJ}) \boldsymbol{\sigma}_I \mathbf{M}_I^{-1} \mathbf{x}_{IJ} \qquad (2\text{-}20)$$

$$\mathbf{T}_J(\mathbf{x}_{JI}) = \omega(\mathbf{x}_{JI}) \boldsymbol{\sigma}_J \mathbf{M}_J^{-1} \mathbf{x}_{JI} \qquad (2\text{-}21)$$

Due to $\omega(\mathbf{x}_{IJ}) = \omega(\mathbf{x}_{JI})$ and $\mathbf{x}_{JI} = \mathbf{x}_{IJ}$, substitute Formula (2-20) and Formula (2-21) into Formula (2-19). The momentum equation can be written as[29]

$$\rho_I \frac{D\mathbf{v}_I}{Dt} = \int_{\mathcal{H}_I} \omega(\mathbf{x}_{IJ})(\boldsymbol{\sigma}_I \mathbf{M}_I^{-1} + \boldsymbol{\sigma}_J \mathbf{M}_J^{-1})\mathbf{x}_{IJ} dV_J + \mathbf{b}_I \qquad (2\text{-}22)$$

The discrete form of the momentum equation is given by

$$\rho_I \frac{D\mathbf{v}_I}{Dt} = \sum_{J=1}^{N} \omega(\mathbf{x}_{IJ})(\boldsymbol{\sigma}_I \mathbf{M}_I^{-1} + \boldsymbol{\sigma}_J \mathbf{M}_J^{-1})\mathbf{x}_{IJ} V_J + \mathbf{b}_I \qquad (2\text{-}23)$$

In the momentum equation, using the $\boldsymbol{\sigma}_I \mathbf{M}_I^{-1} + \boldsymbol{\sigma}_J \mathbf{M}_J^{-1}$ form, the symmetry between the particles can be guaranteed, which guarantees the conservation of linear momentum and the conservation of angular momentum.

In the Cauchy stress tensor expression formula (2-2), the pressure p can be calculated through the Tait equation of state in Equation (2-6). According to the Tait equation of state, it can be obtained that the pressure of the fluid is determined by the density, so a small change in density will cause a large pressure oscillation. Especially for the fluid-structure interaction problems, the pressure instabilities and density oscillations may occur in the fluid particles after a long duration simulation. To avoid this issue, the density filter algorithm [30] is adopted to obtain a stable and smooth pressure field as

$$\rho_I^{\text{new}} = \frac{\sum_J \omega(\mathbf{x}_{IJ}) m_J}{\sum_J \omega(\mathbf{x}_{IJ}) V_J} \qquad (2\text{-}24)$$

where $\rho_I^{\text{new}}$ is the corrected density. Moreover, to reduce computation cost and avoid artificial diffusions, the correction is performed every twenty steps.

The rate of shear strain tensor $\dot{\varepsilon}$ in the proposed ULPH method can be written in a discrete form,

$$\dot{\varepsilon}_I = \left(\left(\sum_{J=1}^{N} \omega(\mathbf{x}_{IJ}) \frac{m_J}{\rho_J} \mathbf{v}_{IJ} \otimes \mathbf{x}_{IJ}\right) \mathbf{M}_I^{-1}\right)^{\text{sym}} - \frac{1}{3}\left(\sum_{J=1}^{N} \omega(\mathbf{x}_{IJ}) \frac{m_J}{\rho_J} \mathbf{v}_{IJ} \cdot (\mathbf{M}_J^{-1} \mathbf{x}_{IJ})\right) \mathbf{I} \qquad (2\text{-}25)$$

To reduce unphysical or numerical oscillations and enhance stability when simulating impact/penetration problems, an artificial viscosity term can be added to the right-hand side of the equation of motion. In this work, the Monaghan[12] type artificial viscosity function is used in the computation. It is modified in [29] to obtain the artificial viscosity formula in the ULPH framework as

$$\Pi_I = \alpha h c_0 \rho_0 \sum_{J=1}^{N} \omega(\mathbf{x}_{IJ}) \pi_{IJ} \mathbf{M}_I^{-1} \mathbf{x}_{IJ} V_J \qquad (2\text{-}26)$$

in which α is the coefficient of the artificial viscosity term. It ranges from 0 to 0.5 depending on the problem. The term $\pi_{IJ}$ is expressed as

$$\pi_{IJ} = \begin{cases} \dfrac{\mathbf{v}_{IJ} \cdot \mathbf{x}_{IJ}}{|\mathbf{x}_{IJ}|^2 + (0.1h)^2}, & \mathbf{v}_{IJ} \cdot \mathbf{x}_{IJ} < 0 \\ 0, & \text{otherwise} \end{cases} \qquad (2\text{-}27)$$

Therefore, the discretized motion governing equation of ULPH after applying artificial viscosity can be rewritten as follows.

$$\rho_I \frac{D\mathbf{v}_I}{Dt} = \sum_{J=1}^{N} \omega(\mathbf{x}_{IJ})(\boldsymbol{\sigma}_I \mathbf{M}_I^{-1} + \boldsymbol{\sigma}_J \mathbf{M}_J^{-1})\mathbf{x}_{IJ} V_J + \Pi_I + \mathbf{b}_I \quad (2\text{-}28)$$

## 2.2. Basic concepts and formulations of non-ordinary state-based Peridynamics

In Peridynamic theory, the research object in the spatial domain R is discretized into a series of Peridynamic particles containing all physical information. Such as position, velocity, density, etc. For every particle $X_J(J=1,2,3,\ldots n_1)$ there is a neighborhood of radius $\delta$ in space, the horizon is denoted as $\mathcal{H}_x$, as shown in Fig. 2. It interacts with every particle $X_J(J=1,2,3,\ldots n_I)$ in its neighborhood, $u$ is the displacement vector of the particle. $\xi_{IJ}$ is the relative position, $\xi_{IJ} = \mathbf{X}_J - \mathbf{X}_I$ and the relative displacement is denoted by $\boldsymbol{\eta}_{IJ}$, $\boldsymbol{\eta}_{IJ} = u[x_J,t] - u[x_I,t]$, as illustrated in Fig. 2.

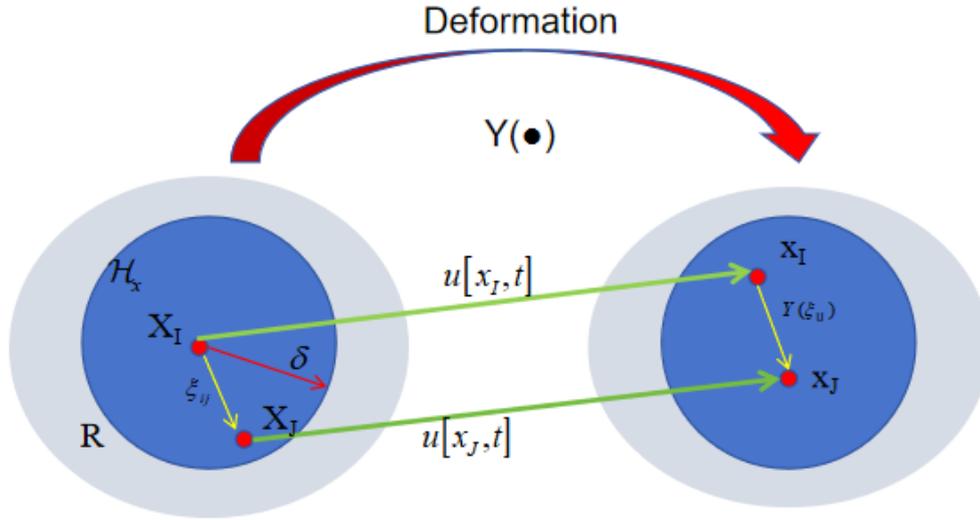

Fig. 2. Schematic diagram of the non-ordinary state-based Peridynamics theory

In continuum mechanics, the equations of motion of a continuum with general dynamic motion are[31]

$$\rho_0 \ddot{\mathbf{u}} = \nabla_{\mathbf{X}} \cdot \mathbf{P}^T + \rho_0 \mathbf{b}, \quad (2\text{-}29)$$

where $\rho_0$ is the current material density. $\ddot{\mathbf{u}}$ is the acceleration of material point. $\nabla_{\mathbf{X}}$ denotes the divergence of the first Piola-Kirchhoff stress P with respect to the reference configuration, and b is the body force. In Peridynamics, the above equation of balance of linear momentum is replaced by a non-local integral equation.

$$\rho_0 \ddot{\mathbf{u}} = \mathbf{L}(\mathbf{x},t) + \rho_0 \mathbf{b} \quad (2\text{-}30)$$

In the NOSB-PD theory, where L(x, t) is a non-local integration of force density vector f(x, x′)

$$\mathbf{L} = \int_V \mathbf{f}(\mathbf{x}^A, \mathbf{x}^B) dV_{\mathbf{x}^B}$$

$$= \int_{\mathcal{H}_{\mathbf{x}_I}} \left[ \mathbf{T}_I(\xi_{IJ}, \mathbf{Y}_I(\xi_{IJ})) - \mathbf{T}_J(\xi_{JI}, \mathbf{Y}_J(\xi_{JI})) \right] dV_{\mathbf{x}^B} \quad (2\text{-}31)$$

where $\mathbf{T}_I(\xi_{IJ}, \mathbf{Y}_I(\xi_{IJ}))$ is the force vector state acted on material point $X_I$ due to material point $X_J$ and likewise for $\mathbf{T}_J(\xi_{JI}, \mathbf{Y}_J(\xi_{JI}))$. Then, the governing equations of motion are rewritten in the NOSB-PD form as[6]

$$\rho_0 \ddot{\mathbf{u}} = \int_{\mathcal{H}_{\mathbf{x}_I}} \left[ \mathbf{T}_I(\xi_{IJ}, \mathbf{Y}_I(\xi_{IJ})) - \mathbf{T}_J(\xi_{JI}, \mathbf{Y}_J(\xi_{JI})) \right] dV_J + \rho_0 \mathbf{b}, \quad (2\text{-}32)$$

where T is the force-vector state. Related to the stress of the first Piola-Kirchhoff We would like to point out that the unit of the force state is N/m$^3$.

$$\underline{\mathbf{T}}[\mathbf{x}_I, t]\langle \xi_{IJ} \rangle = \omega(\xi_{IJ}) \mathbf{P}_{\mathbf{x}_I} \mathbf{K}_{\mathbf{x}_I}^{-1} \langle \xi_{IJ} \rangle \quad (2\text{-}33)$$

$\mathbf{K}_{\mathbf{x}_I}^{-1}$ in Eq. (2-33) represents the inverse of the $\mathbf{K}_{\mathbf{x}_I}$. $\mathbf{K}_{\mathbf{x}_I}$ is the shape tensor of material point $X_I$, that is defined as,

$$\mathbf{K}_{\mathbf{x}_I} = \int_{\mathcal{H}_{\mathbf{x}_I}} \omega(\xi_{IJ}) \xi_{IJ} \otimes \xi_{IJ} dV_{\mathbf{x}_J} \quad (2\text{-}34)$$

$\mathbf{P}_{\mathbf{x}_I}$ in Eq. (2-33) is the first Piola-Kirchhoff stress tensor that can be associated with the Cauchy stress tensor $\sigma$ through,

$$\mathbf{P}_{\mathbf{x}_I} = \mathcal{J} \boldsymbol{\sigma}_{\mathbf{x}_I} \mathbf{F}_{\mathbf{x}_I}^{-T} \quad (2\text{-}35)$$

where $\mathcal{J} = \det \mathbf{F}_{X_I}$, and $\mathbf{F}_{X_I}$ is the nonlocal deformation gradient of particle $X_I$ that is defined as,

$$\mathbf{F}_{\mathbf{x}_I} = \int_{\mathcal{H}_{\mathbf{x}_I}} \left[ \omega(\xi_{IJ}) \underline{\mathbf{Y}} \langle \xi_{IJ} \rangle \otimes \xi_{IJ} \right] dV_{\mathbf{x}_J} \mathbf{K}_I^{-1} \quad (2\text{-}36)$$

### 2.2.1. Failure criterion and short-range

When the relative position between two particles meets certain conditions, the interaction between them will disappear forever, which means the destruction of bond. A bond-breaking indicator $\mu$ [32][33] is introduced to describe the fracture of bonds.

$$\mu(\mathbf{X}_I, t, \xi) = \begin{cases} 1 & s(t, \xi) < s_0 \\ 0 & \text{otherwise} \end{cases} \quad (2\text{-}37)$$

where $s_0$ is the extreme or critical stretch for a given bond, and s is the bond stretch, which is defined by $s = (|\xi + \boldsymbol{\eta}| - |\xi|)/|\xi|$

$$s_0 = \sqrt{5G_0 / (9K\delta_{\mathbf{x}})} \quad (2\text{-}38)$$

where $G_0$ represents energy release rate[34].

In Peridynamics, enabling failure at the bond level is one of its advantages, which leads to an unambiguous local damage $\varphi$ at a material point $X_I$, defined as

$$\varphi(\mathbf{X}_I,t) = 1 - \frac{\int_{H_{\mathbf{x}_I}} \mu(\mathbf{X}_I,t,\xi) \mathrm{d}V_{\mathbf{X}_J}}{\int_{H_{\mathbf{x}_I}} \mathrm{d}V_{\mathbf{X}_J}} \qquad (2\text{-}39)$$

## 2.3. The solid boundary conditions

When the ULPH method is used to numerically simulate problems related to fluid dynamics, there will be boundary types such as free surface boundary, solid wall boundary and periodic boundary. As shown in Figure 2.2, the support domain of fluid particles near the boundary will be truncated by the boundary, which will cause calculation errors and affect the calculation accuracy. Therefore, the study of problems with boundaries requires special treatment at the boundaries.

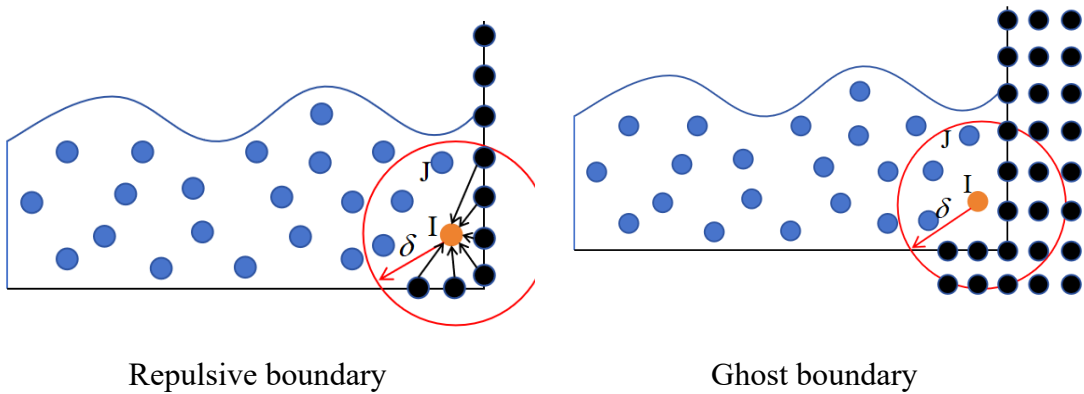

Repulsive boundary            Ghost boundary

Fig. 3.The sketch of different solid boundary treatment methods

For the traditional particle method, there are two main effective methods for solid wall boundary processing: Repulsive boundary method[12], mirror particle boundary[35] and fixed ghost boundary[36]. As shown in Figure 3. Since the Repulsive boundary method has only one layer of boundary particles and does not address the problem of nuclear truncation, it may produce unphysical perturbations to the pressure of the flow field. The mirror particle boundary needs to generate virtual particles dynamically at each step, which reduces the computational efficiency. It is only suitable for regular plane or right-angle boundaries, and it is difficult to determine the position of the mirrored virtual particle for complex boundaries. So in the current work, we adopt the fixed ghost boundary method.

The physical variables of the fixed ghost boundary particles are interpolated from the neighboring fluid particles. The free-slip or no-slip boundary conditions can also be applied to this method [29][37]. For the free-slip solid boundary condition, the viscous force between ghost particles and fluid particles is trivial. For the no-slip solid boundary condition, a virtual velocity $\mathbf{v}_d$ is introduced to implement the interaction of the dummy particles on the fluid particles as follows:

$$\mathbf{v}_s = 2\hat{v}_s - \tilde{\mathbf{v}}_I, \qquad (2\text{-}40)$$

where subscript S represents the ghost particle, $\hat{v}_s$ represents the prescribed velocity of the solid ghost particle $i$ and $\tilde{\mathbf{v}}_I$ represents the interpolation from the neighboring fluid particles as[36]

$$\tilde{\mathbf{v}}_I = \frac{\sum_{J=1}^{N_f} \mathbf{v}_J \omega(\mathbf{x}_{IJ})}{\sum_{J=1}^{N_f} \omega(\mathbf{x}_{IJ})} \quad (2\text{-}41)$$

and $N_f$ refers to the number of neighborhood fluid particles within the horizon of a solid ghost particle I. By substituting Equation (2-40) into the viscous force calculation formula, the viscous force between solid wall particles and fluid particles can be obtained.

The pressure $p_s$ of the solid virtual particle can also be determined by the interpolation of its neighboring fluid particles, and the pressure of the solid wall particle can be regularized by using Shepard kernel [36], so that the final calculation formula of the pressure of the solid wall particle can be obtained as follows.

$$p_s = \frac{\sum_{f=1}^{N_f} p_f \omega(\mathbf{x}_{sf}) - (\mathbf{g} - \mathbf{a}_i) \cdot \sum_{f=1}^{N_f} \rho_f \mathbf{x}_{sf} \omega(\mathbf{x}_{sf})}{\sum_{f=1}^{N_f} \omega(\mathbf{x}_{sf})}, \quad (2\text{-}42)$$

The above equation shows that only the fluid particles in the support domain are considered in the interpolation calculation of the solid wall particle pressure. In the above equation, $\mathbf{a}_i$ is the acceleration of the solid wall boundary[38]

$$\mathbf{a}_I = \frac{D\mathbf{U}}{Dt} + \frac{D\mathbf{\Omega}}{D} \times (\mathbf{x}_I - \mathbf{x}_c) + \mathbf{\Omega} \times (\mathbf{v}_I - \mathbf{U}) \quad (2\text{-}43)$$

U is the translational velocity of the rigid body and $\Omega$ is the rotational velocity. For the fixed solid wall boundary condition, the acceleration of the wall particle is set to $\mathbf{a}_i = 0$; Based on the pressure of the solid particle obtained by interpolation, the density of the solid particle can be evaluated by the equation of state [38]:

$$\rho_I = \frac{p_I}{c_0^2} + \rho_0 \quad (2\text{-}44)$$

Subsequently, the mass of the solid particle can be obtained as

$$m_I = \rho_I V_0 \quad (2\text{-}45)$$

where $V_0$ is the initial volume of solid particle.

### 2.4. Time integration scheme

After discretization of the fluid-structure coupling model, additional solution strategies and update algorithms are needed to meet the requirements of accuracy and stability during the calculation, as well as the computer program implementation of the boundary conditions. This section focuses on the time integration method for explicit dynamical equations, the combined solution strategy for fluid and solid solvers.

Choosing the appropriate time integration method will also greatly affect the running efficiency of the program. For the solid part of PD calculated in this paper, the Velocity and displacement of the solid particle are updated by the Velocity-Verlet time integration method with given boundary conditions and initial conditions:

$$\dot{\mathbf{u}}\left[\mathbf{x}_i, t + \frac{\Delta t}{2}\right] = \dot{\mathbf{u}}[\mathbf{x}_i, t] + \frac{\Delta t}{2} \times \ddot{\mathbf{u}}[\mathbf{x}_i, t]$$

$$\mathbf{u}(\mathbf{x}_i, t + \Delta t) = \mathbf{u}(\mathbf{x}_i, t) + \Delta t \times \dot{\mathbf{u}}\left[\mathbf{x}_i, t + \frac{\Delta t}{2}\right] \quad (2\text{-}46)$$

$$\dot{\mathbf{u}}[\mathbf{x}_i, t + \Delta t] = \dot{\mathbf{u}}\left[\mathbf{x}_i, t + \frac{\Delta t}{2}\right] + \frac{\Delta t}{2} \times \ddot{\mathbf{u}}[\mathbf{x}_i, t + \Delta t]$$

where $\Delta t$ is the time step, $\dot{u}$ and u are the velocity vector and the displacement vector respectively. To maintain the stability and accuracy of the simulation, the size of the time step t should satisfy the CFL condition[38].

Due to the nature of the updated Lagrangian particle hydrodynamics algorithm. In this study, we used the predictor-corrector method for the ULPH time integration of the fluid part, which is divided into two stages:

In the prediction step      In the correction step

$$\begin{cases} \rho_I^{n+\frac{1}{2}} = \rho_I^n + \frac{\Delta t}{2}\left(\frac{d\rho}{dt}\right)^n \\ \mathbf{v}_I^{n+\frac{1}{2}} = \mathbf{v}_I^n + \frac{\Delta t}{2}\left(\frac{d\mathbf{v}}{dt}\right)^n \\ \mathbf{x}_I^{n+\frac{1}{2}} = \mathbf{x}_I^n + \frac{\Delta t}{2}\mathbf{v}_I^{n+\frac{1}{2}} \end{cases} \qquad \begin{cases} \rho_I^{n+1} = \rho_I^n + \Delta t\left(\frac{d\rho}{dt}\right)^{n+\frac{1}{2}} \\ \mathbf{v}_I^{n+1} = \mathbf{v}_I^n + \Delta t\left(\frac{d\mathbf{v}}{dt}\right)^{n+\frac{1}{2}} \\ \mathbf{x}_I^{n+1} = \mathbf{x}_I^n + \Delta t\mathbf{v}_I^{n+1} \end{cases} \quad (2\text{-}47)$$

## 3. NOSB-PD with ULPH coupling scheme

In this study, the method of partition decoupling solution is used to solve the governing equations of the fluid-structure coupling system. For the partition decoupling solution, the fluid and solid parts are solved separately by their own solvers, and the data are exchanged through the coupling interface to meet the coupling conditions. The NOSB-PD theory is utilized in the solid region to describe the material behavior of solids due to its ease of handling damage or rupture processes, while ULPH is used to model fluids. Therefore, a key step is how to deal with the coupling interface in the computational domain of PD-ULPH to guarantee the transfer of force and deformation. Figure 4 illustrates the PD-ULPH coupling scheme based on virtual particles.

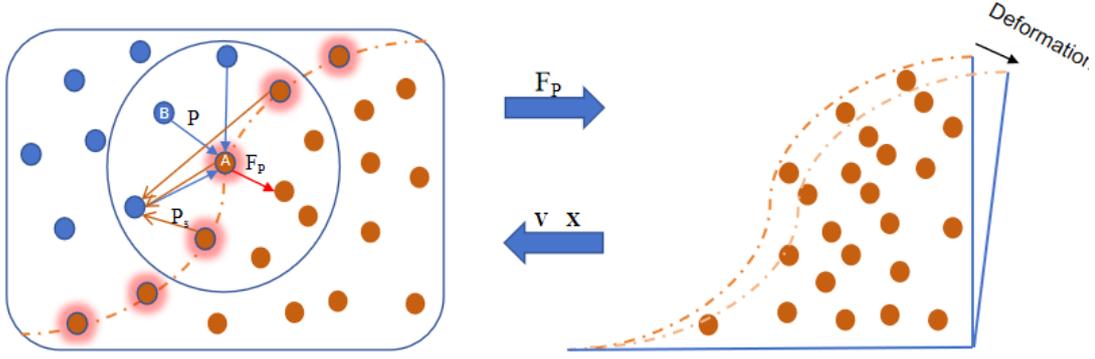

Fig.4 Schematic illustration of the PD-ULPH coupling scheme.

We first consider the force transfer mechanism from ULPH particles to PD particles. Suppose there is a NOSB-PD particle A near the interface whose horizon contains an ULPH particle B. When we compute the force state $\mathbf{T}^A(\xi, \mathbf{Y}(\xi))$ of particle A, the force state of the solid particle is calculated separately, ignoring the fluid particles in the immediate neighborhood. Only the forces of solid particles on solid particles in the support domain are considered.

There are usually two approaches to the force exerted by a fluid particle on a solid particle, a convenient approach is to implement the analysis from the continuum perspective. The forces exerting on solid bodies can be calculated by integrating the stresses along the solid (structures) surface (boundary) [39]. The other is that the pressure of the neighboring fluid particles in the support domain of the solid particle is directly applied to the fluid particle to obtain the force of the fluid acting on the solid[28]. Here we adopt the latter.

$$f_p^B = \sum_{B=1}^{N_f} p_B \frac{\mathbf{x}_{AB}}{|\mathbf{x}_{AB}|} \qquad (2\text{-}48)$$

A similar situation to that mentioned above occurs when a Peridynamic particle A is located inside the support zone of an ULPH particle. When we calculate the interaction force that acts on an ULPH particle but is induced by a Peridynamic particle, we consider the Peridynamic particle as an ULPH particle. Therefore, it participates in the calculation of every conservation law of that ULPH particle. For example, for the ULPH particle B, its linear momentum equation reads as,

$$\rho_B \frac{D\mathbf{v}_B}{Dt} = \sum_{J=1}^{N} \omega(\mathbf{x}_{BA})(\boldsymbol{\sigma}_B \mathbf{M}_B^{-1} + \boldsymbol{\sigma}_A \mathbf{M}_A^{-1})\mathbf{x}_{BA} V_A + \Pi_B + \mathbf{b}_B \Delta x \qquad (2\text{-}49)$$

For Peridynamic particle A which is treated as an ULPH particle, when using equation (2-3) to calculate $\boldsymbol{\sigma}_A$, it is necessary to obtain its pressure $P_B$ and velocity V. Its pressure is interpolated from the ULPH particle in the neighborhood to obtain Equation (2-42), the density and mass are obtained as in Equation (2-44) and Equation (2-45), and the velocity is calculated from Equation (2-40). Therefore, A fixed ghost boundary condition is used to generate interaction between virtual particle A and fluid particle. At the same time, solid particles will provide fluid-solid boundaries for fluid particles, and the pressure interpolated on the solid will generate repulsive forces for fluid particles to prevent particles from penetrating each other.

It is worth mentioning that in this coupling method, the shape tensor of the liquid will become an ill-conditioned matrix due to the influence of the solids in its neighborhood. When inverting the ill-conditioned shape tensor, a small disturbance will cause a large change in the inverse of the shape tensor, which will affect the accuracy of the calculation. So when calculating the shape tensor, the fluid is calculated in the current configuration and the solid in the neighborhood is also in the current configuration, while the solid should remove the influence of the fluid in the neighborhood and be calculated in the initial configuration.

## 4. Validation, application, and discussion

In order to verify the effectiveness of the proposed coupling algorithm and the whole ULPH-NOSBPD framework, this chapter respectively verifies the solid part and fluid part through numerical examples to check the correctness of the solution of solid and fluid, and then compares the correctness and stability of the constructed fluid-structure coupling method through numerical modeling and simulation analysis of fluid-structure coupling problems.

### 4.1. A cantilever beam subjected to concentrated load

In order to verify the effectiveness of the solid solver, we consider the quasi-static problem with a two-dimensional cantilever subjected to a concentrated force, and considering the quasi-static problem, we here adopt the slow loading of the concentrated force on the solid structure.

The initial geometry and Peridynamic model of the beam are shown in Fig. 4, and the configuration parameters are summarized in Table 1. For the present beam, the analytical solution for deflection of the midpoint at the free end can be given as[40].

$$s(L) = -F\left(\frac{L^3}{EI} + \frac{3L}{2GA}\right) \qquad (2\text{-}50)$$

where F is concentrated load, EI and GA are bending and shear stiffness, respectively.

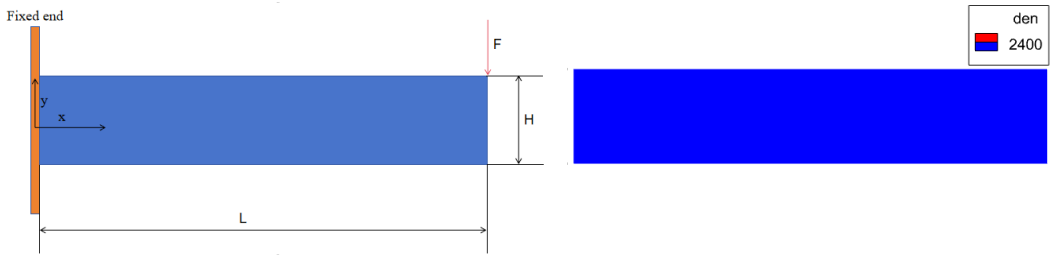

Fig. 4. Initial geometry of the cantilever beam.[40]
(a) Initial geometry model; (b) Peridynamic model.

As shown in Figure 5, the displacement fields calculated by the near-field dynamic model and the finite element calculation model are compared under the concentrated load of the two-dimensional cantilever beam. It can be clearly seen that the displacement fields obtained by the two methods are basically consistent.

Table 1 Physical and numerical parameters for the cantilever beam.

| Parameters | Values |
| --- | --- |
| $L$ | $1m$ |
| $H$ | $0.2m$ |
| Solid density $\rho_s$ | $2400 kg/m^3$ |
| Poisson coefficient $\nu$ | 0.3 |
| Young's Modulus E | $22 GPa$ |
| Particle spacing $\Delta x$ | $0.01m$ |
| Time increment $\Delta t$ | $5 \times 10^{-6} s$ |

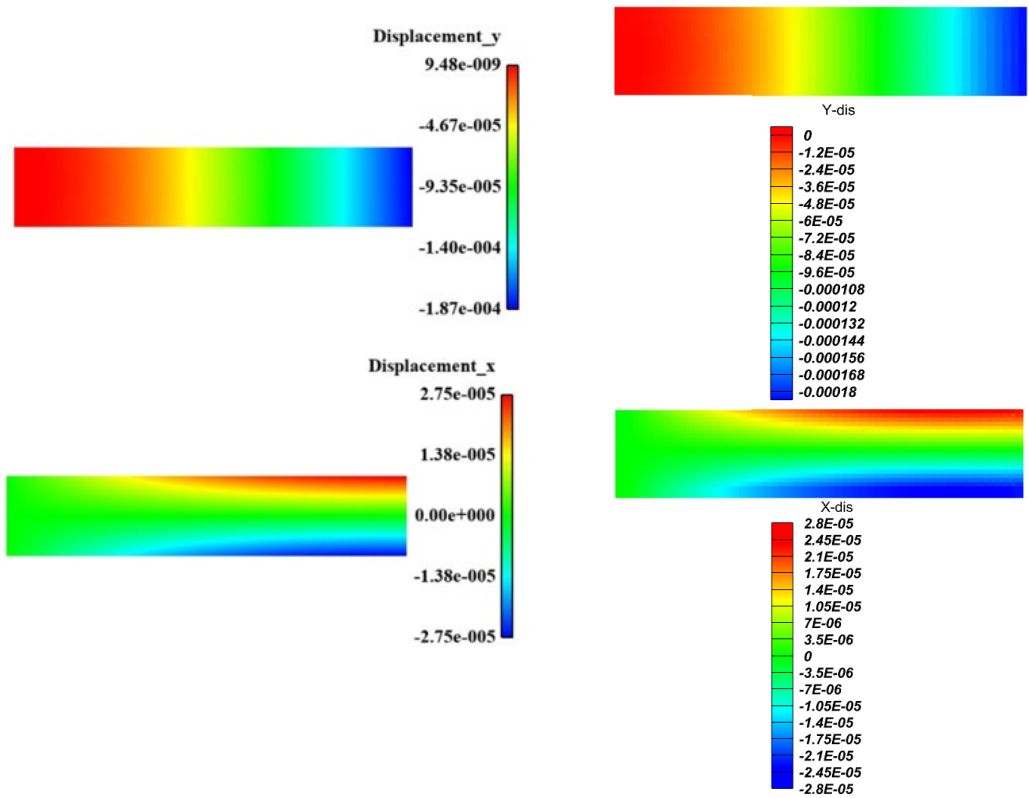

Fig.5 (a) horizontal and vertical displacements given by FEM,
(b) are corresponding results given by peridynamics.

In combination with the above, the NOSB-PD structure solver obtained in this study has the ability to qualitatively and quantitatively simulate the elastic solid problem, and this structure solver will be used in the fluid-structure interaction model later.

## 4.2. Water column collapse problem in a tank

Secondly, we validate our ULPH solver by simulating the famous water column collapse problem in a tank. The length and height of the water column are taken as L= H = 57 mm while the length of tank is given as 4H. The geometric model of the problem is the same as in Martin and Moyce's experiment[27], as shown in Figure 6. Both sides and bottom boundaries are slip-free boundaries, and the relevant parameters are shown in Table 2.

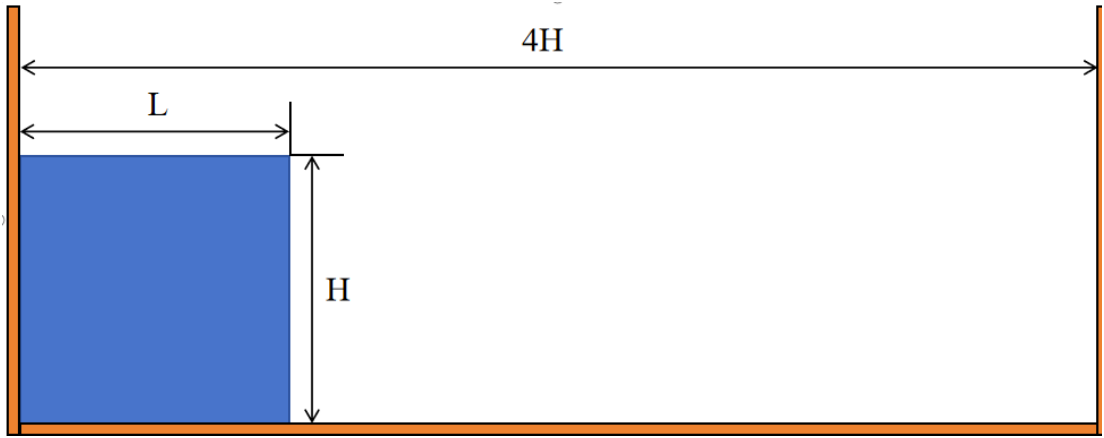

Fig. 6. Initial geometric configuration of the water column collapse problem in a tank

Table 2 Physical and numerical parameters

| Parameters | Values |
| --- | --- |
| L | 0.057 m |
| H | 0.057 m |
| Water density $\rho_w$ | $1000 kg/m^3$ |
| Artificial viscosity coefficient $\alpha$ | 0.02 |
| Reynolds number | 120 |
| Particle spacing $\Delta x$ | $0.001 m$ |
| Time increment $\Delta t$ | $5 \times 10^{-6} s$ |
| Sound speed c | 40m/s |

Figure 7 shows the pressure field contour map every 0.1s for different particle resolutions. It can be seen that for each time, the current ULPH solver is able to accurately capture the pressure gradient distribution and the surface profile of the water before and after hitting the left solid wall boundary. Compared with the results of SPH[41][42], it is also highly consistent. It can be seen that the pressure field and free surface profile of the current ULPH algorithm pressure simulation are in good agreement with the experimental study by Martin et al.

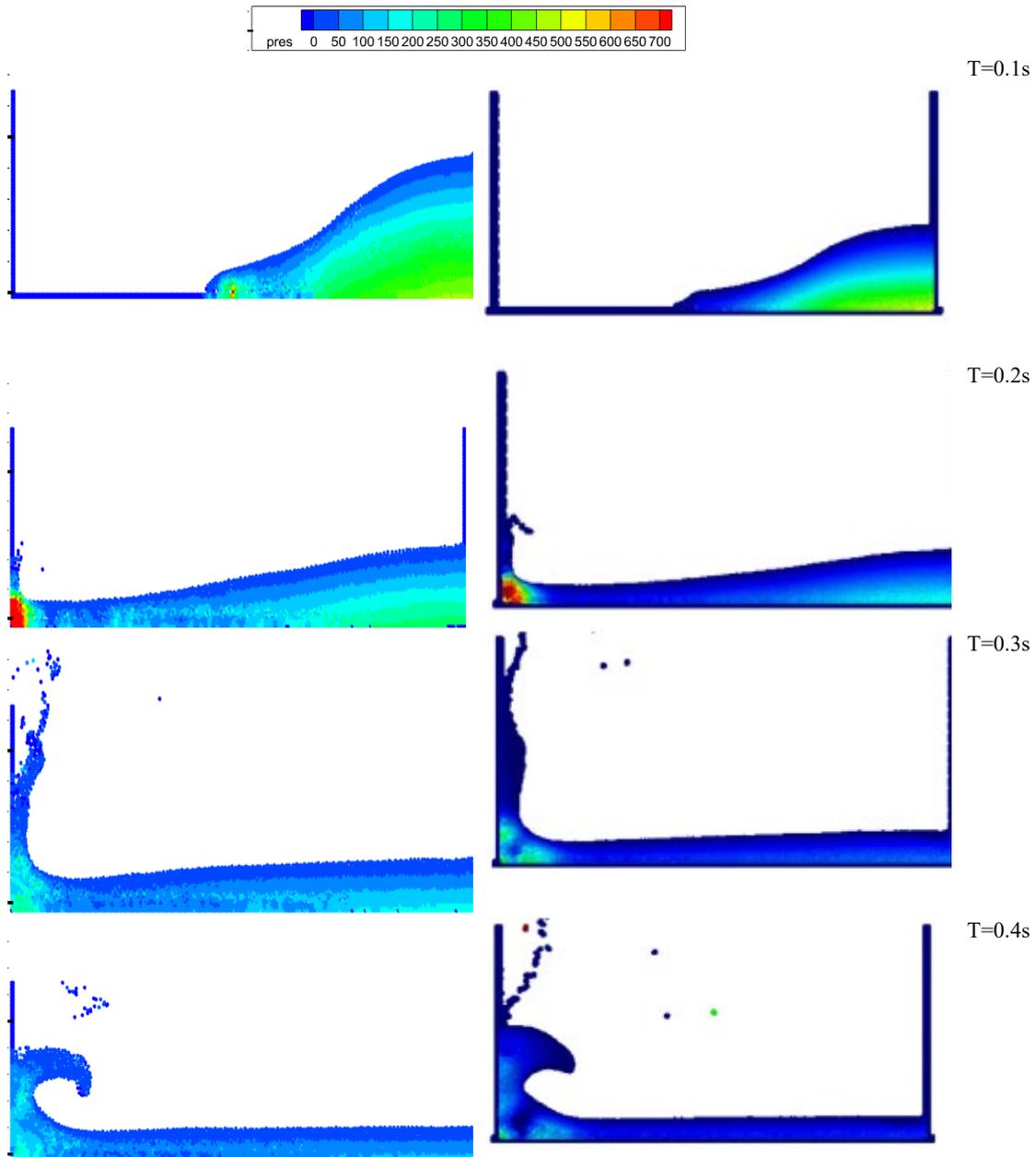

Fig. 7 Contour plot of the pressure field and surface evolution of the fluid by (a)ULPH (b)SPH[41]

### 4.3. A sloshing tank

In this section, the tank sloshing problem is studied based on the ULPH single-phase flow model.

The model parameters of the numerical example are set according to the experiment of Faltinsen et al. [43] The length of the rectangular tank is L = 1.73m, the height is D = 1.15m, and the water depth in the tank is H = 0.5m at the initial time, as shown in Figure 8. At the free surface, a measurement point FS1 is set at 0.05 m from the left wall of the liquid tank, which is used to record the evolution of the water surface height over time. The liquid density in the rectangular tank is $\rho$ =1000kg/m$^3$, and the gravitational acceleration is g=9.81m/s$^2$. The Reynolds number is 120, the speed of sound is 30m/s, and the value of α is 0.1. The rectangular liquid tank is excited by a regular sinusoidal excitation in the horizontal direction (X-axis), and the motion velocity of the rectangular liquid tank[44] is

$$\begin{cases} u(t) = -\dfrac{2\pi}{T} A_0 \sin\left(\dfrac{2\pi}{T} t\right) \\ v(t) = 0 \end{cases} \quad (2\text{-}51)$$

Here, $A_0$=0.032 m is the amplitude and T = 1.875 s is the period. The initial particle spacing of the computational domain is Δx=0.01 m. This example is simulated for a total of 6 s, and the reference sound speed is set to $c_0$=40 m/s.

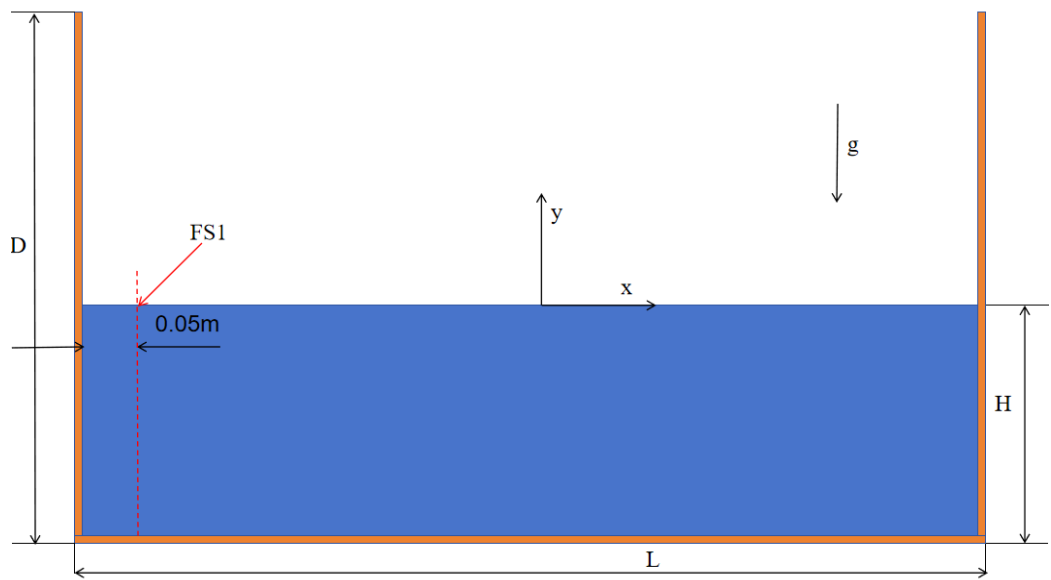

Fig. 8 Sketch of initial setup of the sloshing tank

Figure 9. shows the evolution of the sloshing liquid level and the distribution of the pressure field in the tank at different times under horizontal excitation. It can be observed from the figure that under the action of periodic external excitation, the tank moves back and forth in the horizontal direction, causing the water in the tank to move back and forth and produce large liquid surface deformation. The pressure field of water in the figure is smooth, without pressure oscillation, and the distribution of particles at the liquid surface is continuous, without non-physical gaps. Therefore, the stability and accuracy of the ULPH fluid model in the simulation of large deformation free surface flow problems can be concluded.

Figure 3.7 shows the evolution of the water surface height at the measured point over time, and compares it with the experimental results of Faltinsen[43] and the simulation results of Yan[45]. Comparing the ULPH results with the experimentally measured data, it can be seen that the ULPH measured results are in good agreement with the experimental data.

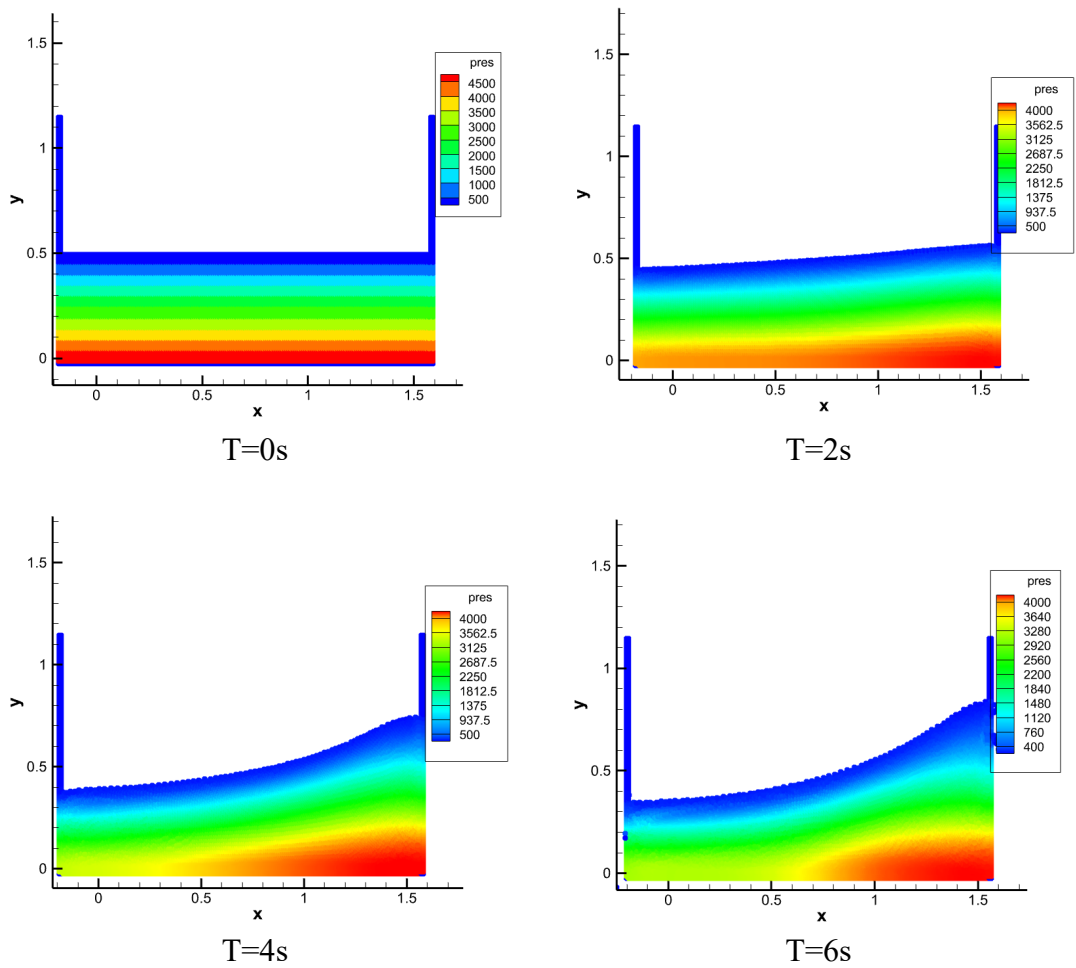

Fig. 9.The development of the free surface and pressure field distribution of the rectangular tank sloshing at different times under horizontal excitation

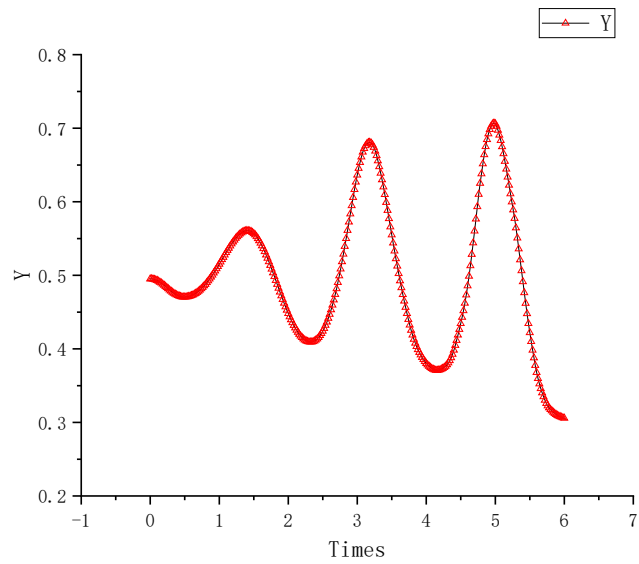

Fig. 9 The change of the water surface height with time at the measurement point FS1

## 4.4. Breaking dam impacting on an elastic plate

To demonstrate the effectiveness of the FSI framework proposed in this work for violent free-surface flows interacting with deformable structures, dam-break flows impacting elastic plates have been modeled, which has been extensively simulated as an appropriate benchmark to validate numerical models of FSI problems[42][46][47].

Figure 10. shows the initial appearance of this model, with water of width L and height 2L initially located on the left and bottom walls. The distance between the two vertical walls is 4L. An elastic baffle is fixed at the bottom end at distance L from the water column, and the top of the baffle is free and the bottom is fixed. Under the force of gravity, the water column collapses rapidly and rushes towards the right boundary. Strong FSI occurs when the flow front collides with the baffle. In this case, baffles are considered as ideal elastomers. The material parameters and fluid parameters of the baffle are shown in the Table.3

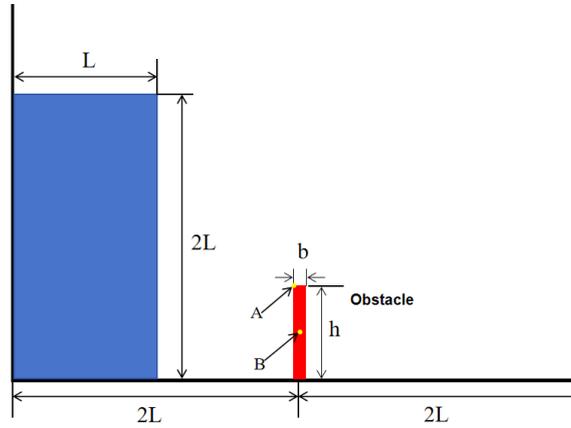

Fig. 10. Geometric configuration of the water impact on an elastic plate.

Table.3 Physical and numerical parameters

| Parameters | Values |
| --- | --- |
| L | $0.146m$ |
| h | $0.08m$ |
| a | $0.012m$ |
| Solid density $\rho$ | $2500 kg/m^3$ |
| Poisson coefficient $\upsilon$ | 0.0 |
| Young's Modulus E | $1MPa$ |
| Particle spacing $\Delta x$ | $0.002m$ |
| Time increment $\Delta t$ | $5\times 10^{-6} s$ |
| Sound speed | 60m/s |
| Reynolds number | *120* |
| Artificial viscosity coefficient $\alpha$ | *0.1* |

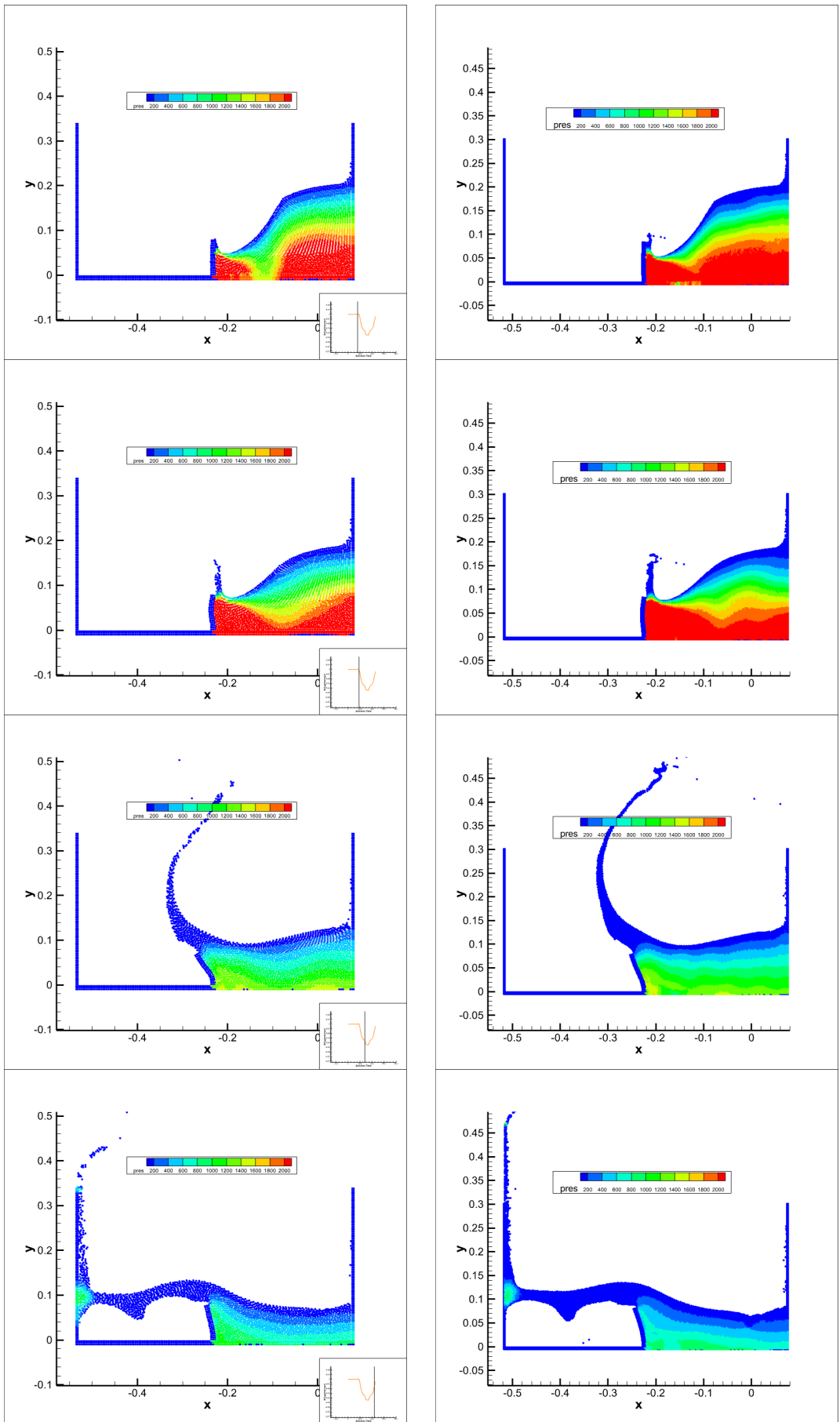

Fig.11 dam break flow impacting on baffle based on different numerical models.

Figure 11. shows the simulation of the dam-break flow impacting the baffles and the pressure distribution of the fluid and the deformation of the structure obtained using other numerical models. The results show that the coupled model successfully reproduces the pressure field and structural deformation near the fluid-solid interface. At the same time, the flow state, pressure distribution and structural deformation are consistent with the results of SPH-PD[47][46][42] simulation and FPM-SPH [48]simulation.

Figure 12 shows the evolution of the horizontal displacement of the free end of the baffle, point A in Figure 10. It can be seen that after the wave front of the burst dam reaches the elastic plate, the pressure at the lower part of the fluid-elastic plate interaction area increases rapidly, making the elastic plate deflector to the maximum value. At this stage (0.15s-0.23s), the free end of the elastic plate undergoes high-speed deformation. As the fluid moves on the plate, the deflection of the plate decreases as the fluid pressure on the plate decreases, and a rebound will follow. The water column flowing down the baffle eventually hits the vertical wall on the right side, where the impact on the hard wall produces a violent splash, undergoes a regional instantaneous pressure increase, and finally gradually falls due to gravity.

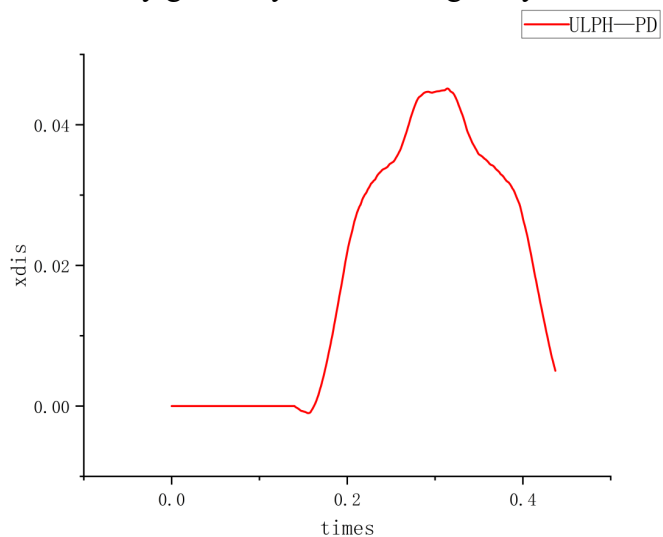

Fig.12. Comparison of the predicted evolution of the horizontal displacement of point A based on different numerical models.

To analyze the fracture behavior of the structure under the impact of dam-break flow, the baffle in Fig. 10 was use the Drucker-Prager[50] model to simulate the fluid-structure interaction and consider the damage problem of the solid in the solid constitution in the previous section with a critical bond stretch ratio of s0 = 0.069. The other parameters involved in the simulation were the same as [51]. Fig. 13 presents simulation snapshots of the dam-break flow propagation and the brittle fracturing of the baffle. The pressure distribution in the water, and the displacement in x direction field in the water near the baffle were obtained.

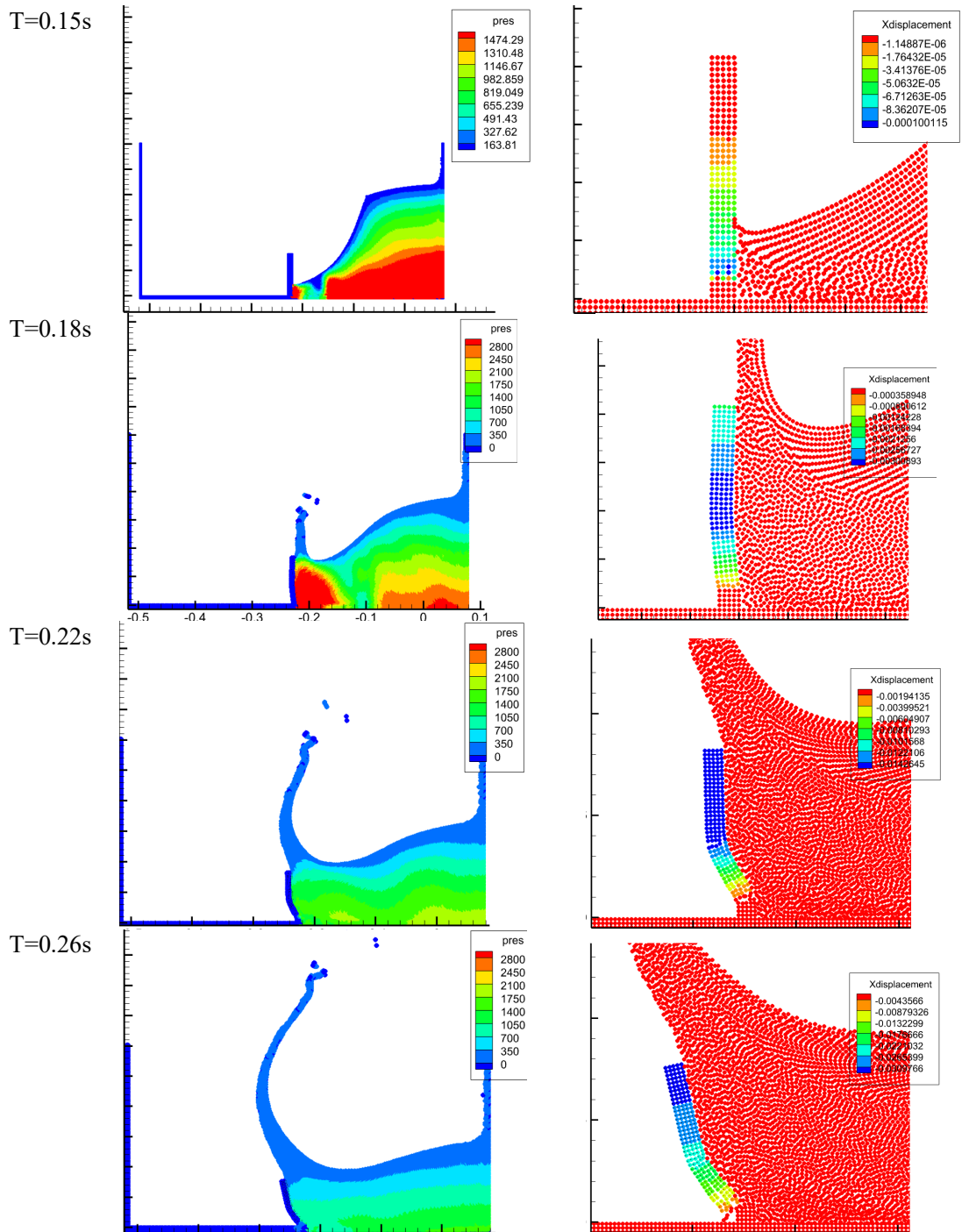

Fig.13 Snapshot of pressure, displacement in x direction by proposed method.

### 4.5. Interaction of a dam-break wave with an elastic sluice gate

The second FSI validation example for the ULPH-PD coupled model is the interaction between the dam-break flow and the elastic gate which was first investigated experimentally and numerically by Yilmaz et al. The initial geometry of the model is shown in the figure. The height of the water column is 0.2m and the width is 0.5m, and the water column is initially in a static state. There is an elastic gate 0.3m away from the water column, the length of the elastic gate is 0.125m and the width is 0.007m. The material parameters and fluid specific parameters of the elastic gate are shown in the Table 4.

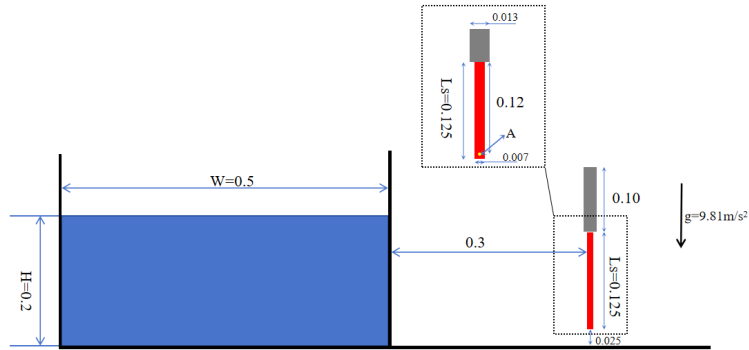

Fig.13. Geometric configuration of the interaction of a dam-break wave with an elastic sluice gate

Table 4 Physical and numerical parameters

| Parameters | Values |
| --- | --- |
| Solid density ρ | $1250 kg/m^3$ |
| Poisson coefficient $\upsilon$ | 0.4 |
| Young's Modulus E | $4MPa$ |
| Particle spacing $\Delta x$ | $0.004m$ |
| Time increment $\Delta t$ | $5 \times 10^{-6}s$ |
| Sound speed | 60m/s |
| Reynolds number coefficient $\alpha$ | 120 |

　　Figure 13. shows the comparison of the results of the PD-ULPH coupled model of this problem with the experimental results and the simulation results of other meshless methods. It can be seen that the PD-ULPH model can well predict the position of the water free surface and the deformation of the elastic gate, and the pressure field and horizontal displacement field are also smooth. At time t =0 s, the solid wall on the right is released and the water column begins to collapse. After 0.2s, the fluid starts to impact the elastic gate, and the outlet formed by the hydraulic pressure of the elastic gate gradually increases at first. When it reaches the maximum value, the outlet will gradually decrease with the decrease of the water pressure.

　　Figure14 shows the horizontal displacement comparison at the measurement point A. The current coupled simulation results are in good agreement with the experimental measurement data [49] and the numerical simulation results of Yilmaz[49] et al.

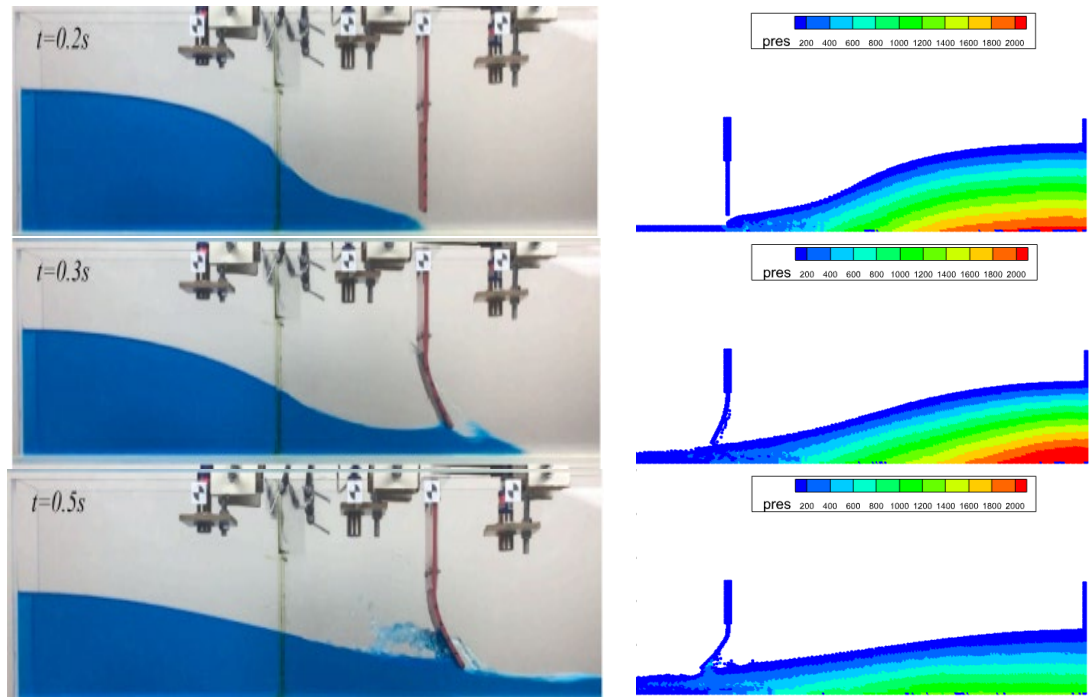

Fig.13. Comparision frames of the experimental and numerical results at various time steps.

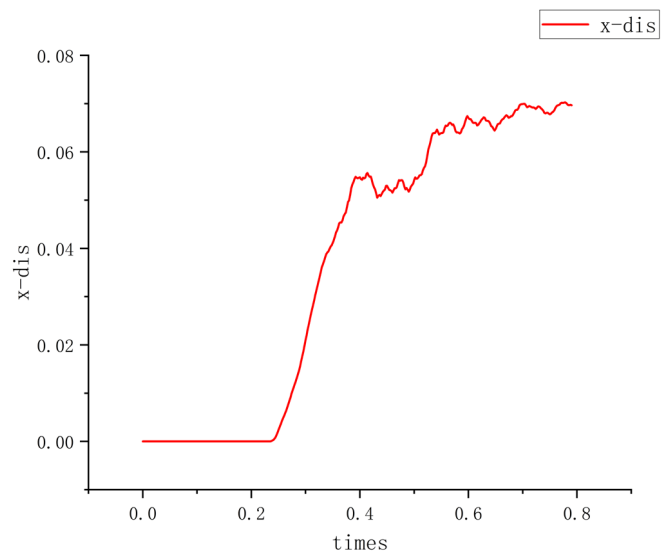

Fig.14. Time histories of horizontal displacements at measurement point A